\documentclass[%
 reprint,
showpacs,preprintnumbers,
 amsmath,amssymb,
prb,
superscriptaddress
]{revtex4-2}

\usepackage{graphicx}
\usepackage{dcolumn}
\usepackage{bm}
\usepackage{physics}
\usepackage{comment}
\usepackage{algpseudocode}
\graphicspath{{./figures/}}

\usepackage{color}

\begin{document}


\title{Electron Excitation Probability in Dielectrics under Two-color Intense Laser Fields}

\author{Mizuki Tani}
\email{tani.mizuki@qst.go.jp}
\affiliation{%
 Kansai Institute for Photon Science, National Institutes for Quantum Science and Technology (QST), 8-1-7 Umemidai, Kizugawa, Kyoto 619-0215, Japan
}
\affiliation{%
 Photon Science Center, Graduate School of Engineering, The University of Tokyo, 7-3-1 Hongo, Bunkyo-ku, Tokyo 113-8656, Japan
}
\affiliation{%
 Department of Nuclear Engineering and Management, Graduate School of Engineering, The University of Tokyo,7-3-1 Hongo, Bunkyo-ku, Tokyo 113-8656, Japan
}

\author{Kenichi L. Ishikawa}
\email{ishiken@n.t.u-tokyo.ac.jp}
\affiliation{%
 Photon Science Center, Graduate School of Engineering, The University of Tokyo, 7-3-1 Hongo, Bunkyo-ku, Tokyo 113-8656, Japan
}
\affiliation{%
 Department of Nuclear Engineering and Management, Graduate School of Engineering, The University of Tokyo,7-3-1 Hongo, Bunkyo-ku, Tokyo 113-8656, Japan
}%
\affiliation{%
 Research Institute for Photon Science and Laser Technology, The University of Tokyo, 7-3-1 Hongo, Bunkyo-ku, Tokyo 113-0033, Japan
}
\affiliation{%
Institute for Attosecond Laser Facility, The University of Tokyo, 7-3-1 Hongo, Bunkyo-ku, Tokyo 113-0033, Japan
}

\author{Tomohito Otobe}%
\email{otobe.tomohito@qst.go.jp}
\affiliation{%
 Kansai Institute for Photon Science, National Institutes for Quantum Science and Technology (QST), 8-1-7 Umemidai, Kizugawa, Kyoto 619-0215, Japan
}
\affiliation{%
 Photon Science Center, Graduate School of Engineering, The University of Tokyo, 7-3-1 Hongo, Bunkyo-ku, Tokyo 113-8656, Japan
}

\date{\today}

\begin{abstract}
Two-color laser fields offer significantly enhanced control over electron excitation dynamics under ultrashort intense laser pulses compared to monochromatic fields. However, their strong nonlinearity necessitates computationally expensive first-principles calculations to accurately predict ionization dynamics. To overcome this challenge, we derive an analytical expression for the ionization rate in dielectrics subjected to intense two-color laser fields, refining the theoretical framework introduced in JPSJ {\bf 88}, 024706 (2019). By benchmarking our formula against first-principles calculations based on time-dependent density functional theory (TDDFT) for $\alpha$-quartz, we demonstrate that our model captures the essential physics of ionization dynamics with remarkable qualitative accuracy, despite employing certain approximations. This analytical approach not only provides deeper physical insight but also offers a computationally efficient alternative for predicting strong-field interactions in dielectrics.
\end{abstract}

\pacs{Valid PACS appear here}
\maketitle


\section{\label{sec:introduction}INTRODUCTION}
Electron excitation in dielectrics plays a fundamental role in strong laser-matter interactions \cite{Stuart1995-cl}. Depending on the laser intensity, the excitation of electrons from the valence band into the conduction band underlies a wide variety of nonlinear phenomena. This process gives rise to high-order harmonic generation (HHG) \cite{Luu2015-tl} at relatively moderate intensities below the damage threshold, drives the formation of laser-induced periodic surface structures (LIPSS) near the damage threshold \cite{Temple1981-pn}, and, at even higher intensities with femtosecond pulses, enables precision microfabrication with minimal thermal damage \cite{Gattass2008-ru}.
Each of these phenomena is initiated by the dynamics of laser-driven electron excitation, which thus attracts significant interest in both fundamental science and advanced technological applications \cite{Sugioka2014-fh}.

Optical nonlinear electron excitation processes can be numerically simulated, e.g., by the semiconductor Bloch equations, the time-dependent density matrix method \cite{Haug1984-ik,Golde2008-yi,Kaneshima2018-di,Yue2022-rn}, the time-dependent density functional theory (TDDFT) \cite{Iwata2001-hu,Otobe2008-yz,Noda2019-px,Tancogne-Dejean2020-bm}, and its semiclassical approximation, i.e., Vlasov equation \cite{Fennel2004-vx,Kohn2008-hz,Fennel2010-zd,Tani2021-vf}.
These methods can be flexibly applied to different materials and practically arbitrary pulse shapes, and the results can be directly compared with experiments.

In parallel, analytical theories are powerful tools for gaining physical insight into various physical mechanisms such as multiphoton absorption and tunneling.
In his seminal work, Keldysh proposed an analytical solution for the probability of electron excitation in both atoms and semiconductors \cite{Keldysh1965-pj}, which has been widely used to understand strong field phenomena \cite{Lewenstein1994-ss, Golin2014-eq, Kozak2019-vj}. 
The Keldysh theory, originally formulated in the length gauge and for linear polarization, has since been extended to multiple directions; it has been applied to atoms under circularly polarized laser fields \cite{Faisal1973-gn}, reformulated in the velocity gauge with further extension to solids \cite{Jones1977-vs}, and adapted to solids under circularly and elliptically polarized laser fields \cite{Reiss1980-vg,Otobe2019-cv,Venkat2022-mn}. The periodicity effect in the solid was incorporated \cite{Gruzdev2007-ow}, and the theory was even related to the Schwinger mechanism \cite{Taya2021-iu}.

While all of these previous analytical theories have, to the best of our knowledge, considered single-color irradiation, the use of two-color laser pulses has recently been attracting considerable attention in both experiment and theory.
There have been studies reporting that synthesized dual-color laser pulse pairs achieve highly efficient laser drilling compared to single-color irradiation \cite{Sugioka1993-hg,Yu2013-aw,Gedvilas2017-kl,Zhukov2024-tq,Yao2025-sw}.
Two-color pulses can dramatically enhance the HHG intensity in atomic \cite{Ishikawa2003-qx,Takahashi2007-rd} and solid targets \cite{Song2020-hk}, as well as increase the electron excitation rate in band-gap materials \cite{Duchateau2022-ba,Tani2022-yy}.
These studies have clearly shown that the outcomes of dual-color irradiation are often essentially distinct from simply combining the effects of individual color excitations or from those typically observed in coherent control scenarios \cite{Sheehy1995-zf,Endo2017-no,Di-Fraia2019-sz,You2020-tv,Orimo2021-lw,Gryzlova2022-bh,Ishikawa2023-hg}.

Therefore, a new development of an analytical theory of electron excitation probability in dielectrics tailored to two-color excitation is in order, which is the objective of this study.
Using the Houston basis, we derive an analytical expression for the electron excitation probability in a parabolic two-band system under strong two-color laser fields.
We compare, for $\alpha$-quartz, the results of our analytical formula with those of first-principles calculations using time-dependent density functional theory (TDDFT) \cite{Runge1984-xs,Yabana1996-vi,Noda2019-px} and find a good qualitative agreement.

This paper is organized as follows; methods for analytical derivation of formula and TDDFT calculation are described in Sec.~\ref{sec:formalism}, their numerical experiments are shown in Sec.~\ref{sec:results}, and concluded in Sec.~\ref{sec:conclusions}.
We use the atomic units  (a.u.) unless states otherwise, i.e. $e=m=\hbar=1$ where $e$, $m$, and $\hbar$ denote elemental charge, electron mass, and Dirac constant, respectively.

\section{\label{sec:formalism} Theoretical formulation}

\subsection{\label{subsec:houston}Electron dynamics in the Houston basis}
Within the independent-electron approximation and the dipole approximation, we start from the time-dependent Schr\"odinger equation in the velocity gauge under vector potential $\mathbf{A}(t)$ related to laser electric field $\mathbf{E}(t)$ by $\mathbf{E}(t)=-\dot{\mathbf{A}}(t)$,
\begin{equation}
    i\frac{\partial}{\partial t} u_{n,\mathbf{k}}(\mathbf{r},t) = \left[ \frac{1}{2}(\mathbf{p}+\mathbf{k}+\mathbf{A}(t))^2 + V(\mathbf{r}) \right]u_{n,\mathbf{k}}(\mathbf{r},t), \label{tdse}
\end{equation}
where $u_{n,\mathbf{k}}(\mathbf{r},t)$ represents the periodic part of the time-dependent wave function of the electron that initially lies in the band $n$ with a wave vector $\mathbf{k}$,
$\mathbf{p}$ the canonical momentum operator,
and $V(\mathbf{r})$ the local potential.

The periodic part of the ground state wave function $u^{\mathrm{G}}_{n,\mathbf{k}}(\mathbf{r})\equiv u_{n,\mathbf{k}}(\mathbf{r},0)$ satisfies the static Schrödinger equation:
\begin{equation}
\epsilon^{\mathrm{G}}_{n,\mathbf{k}} u^{\mathrm{G}}_{n,\mathbf{k}}(\mathbf{r}) = \left[ \frac{1}{2}(\mathbf{p}+\mathbf{k})^2 + V(\mathbf{r}) \right]u^{\mathrm{G}}_{n,\mathbf{k}}(\mathbf{r}), \label{se}
\end{equation}
where $\epsilon^{\mathrm{G}}_{n,\mathbf{k}}$ is the eigen energy.
Using the Houston function defined as,
\begin{equation}
    w_{n,\mathbf{k}}(\mathbf{r},t) = u^{\mathrm{G}}_{n,\mathbf{k}+\mathbf{A}(t)}(\mathbf{r}) \exp(-i\int^t \epsilon^{\mathrm{G}}_{n,\mathbf{k}+\mathbf{A}(t')}dt'), \label{houston}
\end{equation}
let us expand the time-dependent orbital as,
\begin{equation}
    u_{n,\mathbf{k}}(\mathbf{r},t) = \sum_{n'} C_{nn'}^{\mathbf{k}}(t) w_{n',\mathbf{k}}(\mathbf{r},t). \label{expansion}
\end{equation}
Inserting Eq.~(\ref{expansion}) into Eq.~(\ref{tdse}) and using Eqs.~(\ref{se}) and (\ref{houston}),
we obtain the equation of motion for 
the expansion coefficient $C_{nn'}^{\mathbf{k}}(t)$ as,
\onecolumngrid
\begin{gather}
    i\frac{\partial}{\partial t}C_{nn'}^{\mathbf{k}}(t)
    = -i\frac{d\mathbf{A}}{dt} \sum_{n"}C_{nn"}^{\mathbf{k}}(t) \braket{u^{\mathrm{G}}_{n',\mathbf{k}+\mathbf{A}(t)}(\mathbf{r})}{\frac{\partial}{\partial \mathbf{k}}u^{\mathrm{G}}_{n",\mathbf{k}+\mathbf{A}(t)}(\mathbf{r})} \exp(i\int^t \left( \epsilon^{\mathrm{G}}_{n',\mathbf{k}+\mathbf{A}(t')}-\epsilon^{\mathrm{G}}_{n",\mathbf{k}+\mathbf{A}(t')} \right) dt') \nonumber \\
    = \frac{d\mathbf{A}}{dt} \sum_{n"}C_{nn"}^{\mathbf{k}}(t) \frac{
    \bra{u^{\mathrm{G}}_{n',\mathbf{k}+\mathbf{A}(t)}(\mathbf{r})}
    \mathbf{p}
    \ket{{u^{\mathrm{G}}_{n",\mathbf{k}+\mathbf{A}(t)}(\mathbf{r})}}
    }
    {\epsilon^{\mathrm{G}}_{n',\mathbf{k}+\mathbf{A}(t)}-\epsilon^{\mathrm{G}}_{n",\mathbf{k}+\mathbf{A}(t)}} \exp(i\int^t \left( \epsilon^{\mathrm{G}}_{n',\mathbf{k}+\mathbf{A}(t')}-\epsilon^{\mathrm{G}}_{n",\mathbf{k}+\mathbf{A}(t')}  \right)dt'). \label{Cnnp}
\end{gather}
\twocolumngrid
\subsection{\label{subsec:probability} Two-band system}
Equation (\ref{Cnnp}) does not explicitly contain $V(\mathbf{r})$, but the dynamics is mediated by transition matrix elements. 
Thus, instead of specifying the form of $V(\mathbf{r})$,
we consider a two-band system consisting of a conduction band ($c$) and a valence band ($v$) with the energy gap $\Delta^{\mathbf{k}}_{\mathrm{vc}}$.
We assume that the conduction band is initially unoccupied and that the population of the valence band remains nearly unity, i.e., $|C_{\mathrm{vv}}^{\mathbf{k}}(t)| \sim 1 \gg |C_{\mathrm{vc}}^{\mathbf{k}}(t)|$.
Under this assumption, the transition amplitude can be approximated as:
\begin{equation}
    C_{\mathrm{vc}}^{\mathbf{k}}(t) \sim -i\int^t d\tau \left[ \mathbf{A}(\tau) \cdot \mathbf{P}_{\mathrm{vc}}^{\mathbf{k+A}(\tau)} \exp(i\int^\tau \Delta^{\mathbf{k}+\mathbf{A}(t')}_{\mathrm{vc}} dt') \right], \label{Cvc}
\end{equation}
where the momentum matrix element is defined as $\mathbf{P}_{\mathrm{vc}}^{\mathbf{k}} = \mel{u^{\mathrm{G}}_{\mathrm{c},\mathbf{k}}}{\mathbf{p}}{u^{\mathrm{G}}_{\mathrm{v},\mathbf{k}}}$.
This expression represents the time evolution of an electron transitioning between bands under the influence of the external laser field.
To derive Eq.~(\ref{Cvc}), we have performed a partial integration with respect to time, neglected the oscillatory terms not related to the number of excited electrons and terms including $\Delta^{\mathbf{k}+\mathbf{A(t)}}_{\mathrm{vc}}$ which will disappear in the long time limit (see Eq.~(\ref{eq:limit})).

By taking the integral Eq.~(\ref{Cvc}) over a sufficiently long time interval $t \in [-T,T]\,(T>0)$,
\begin{equation}
    \Tilde{C}_{\mathrm{vc}}^{\mathbf{k}} = -i \int_{-T}^T dt \left[ \mathbf{A}(t) \cdot \mathbf{P}_{\mathrm{vc}}^{\mathbf{k+A}(t)} \exp(i\int^t \Delta^{\mathbf{k}+\mathbf{A}(t')}_{\mathrm{vc}} dt') \right], \label{Cint}
\end{equation}
we can express the $k$-resolved transition probability per unit time as,
\begin{equation}
    \tau_{\mathbf{k}} = \frac{|\Tilde{C}_{\mathrm{vc}}^{\mathbf{k}}|^2}{2T}.
\end{equation}
Under a continuous laser field, the total transition probability $W$ per unit time and volume can be defined as,
\begin{equation}
\label{eq:limit}
    W = \lim_{T\to\infty} \frac{2}{8\pi^3}\int_{\mathrm{BZ}} \tau_{\mathbf{k}} d\mathbf{k}.
\end{equation}

\subsection{\label{subsec:two-band}Parabolic two-band system}
We further assume a parabolic two-band system whose energy gap $\Delta^{\mathbf{k}}_{\mathrm{vc}}$ is given by,
\begin{equation}
    \Delta^{\mathbf{k}}_{\mathrm{vc}} \equiv \epsilon^{\mathrm{G}}_{\mathrm{c},\mathbf{k}}-\epsilon^{\mathrm{G}}_{\mathrm{v},\mathbf{k}} = E_{\mathrm{g}} + \frac{\mathbf{k}^2}{2\mu}.
\end{equation}
Here, $E_{\mathrm{g}}$ is the bandgap, and $\mu$ is the reduced mass of the system.
The phase factor in Eq.~(\ref{Cint}) is expressed as,
\begin{gather}
    \exp(i\int^t \Delta^{\mathbf{k}+\mathbf{A}(t')}_{\mathrm{vc}} dt') \notag \\
    = \exp(i\int^t \left[ \frac{k^2}{2\mu} + \frac{\mathbf{k} \cdot \mathbf{A}}{\mu} + \frac{|\mathbf{A}|^2}{2\mu} + E_g \right] dt'). \label{phase}
\end{gather}

Let us specifically consider a collinear two-color laser field, composed of the fundamental frequency $\omega$ and its second harmonic component $2\omega$, whose vector potential is given by,
\begin{equation}
    \mathbf{A}(t) = (0,0,A_1 \cos{\omega t} + A_2 \cos{(2\omega t + \phi)}), \label{vector_potential}
\end{equation}
where $\phi$ denotes the $\omega$-$2\omega$ relative phase, and $A_1$ and $A_2$ the amplitudes of the $\omega$ and $2\omega$ components, respectively.

\subsubsection{Zero relative phase $\phi=0$}
First, let us examine the $\phi=0$ case.
Equation~(\ref{phase}) is transformed into,
\begin{gather}
    \exp(i\int^t \Delta^{\mathbf{k}+\mathbf{A}(t')}_{\mathrm{vc}} dt') \notag \\
    = \sum_l \exp(i\left[\frac{k^2}{2\mu} + U_p + E_g + l\omega \right] t) J_l(\eta_1,\eta_2,\eta_3,\eta_4), \label{phase_exp}
\end{gather}
where $U_p=(A_1^2+A_2^2)/4\mu$ is the ponderomotive energy, $J_l(\eta_1,\eta_2,\eta_3,\eta_4)$ stands for the $l$-th order 4-variable Bessel function \cite{Dattoli1992-nz}, defined by,
\begin{gather}
    J_l(\eta_1,\eta_2,\eta_3,\eta_4) \notag \\
    = \sum_{l_2,l_3,l_4} J_{l-2l_2-3l_3-4l_4}(\eta_1)J_{l_2}(\eta_2)J_{l_3}(\eta_3)J_{l_4}(\eta_4),
\end{gather}
with $J_l(\eta)$ being the $l$-th order Bessel function, and,
\begin{gather}
    \eta_1 = \frac{kA_1\cos{\theta}}{\mu\omega}+\frac{A_1A_2}{2\mu\omega},
    \eta_2 = \frac{kA_2\cos{\theta}}{2\omega\mu}+\frac{A_1^2}{8\omega\mu}, \notag \\
    \eta_3 = \frac{A_1A_2}{6\omega\mu},
    \eta_4 = \frac{A_2^2}{16\omega\mu}.
\end{gather}
It should be noticed that the 4-variable Bessel function is reduced to the generalized Bessel function when $A_1=0$ or $A_2=0$,
\begin{gather}
    J_l(0,\eta_2,0,\eta_4) = \sum_{l_4} J_{l-2l_4}(\eta_2)J_{l_4}(\eta_4),
    \\
    J_l(\eta_1,\eta_2,0,0) = \sum_{l_2} J_{l-2l_2}(\eta_1)J_{l_2}(\eta_2),
\end{gather}
which appear in the excitation probability under a linearly polarized monochromatic laser field \cite{Otobe2019-cv}.
Then, $\Tilde{C}_{\mathrm{vc}}^{\mathbf{k}}$ is written as,
\begin{gather}
    \Tilde{C}_{\mathrm{vc}}^{\mathbf{k}} = -i |P_{\mathrm{vc}}| \int_{-T}^T dt \left[ (A_1 \frac{e^{i\omega t}+e^{-i\omega t}}{2} + A_2 \frac{e^{2i\omega t}+e^{-2i\omega t}}{2} \right. \notag \\
    \left. \sum_l e^{i\xi_l t} J_l(\eta_1,\eta_2,\eta_3,\eta_4) \right] \notag \\
    = -\frac{i|P_{\mathrm{vc}}|}{2} \int_{-T}^T dt \sum_l \left[ (A_1 (e^{i\xi_{l+1} t}+e^{i\xi_{l-1} t}) \right. \notag \\
    \left. + A_2 (e^{i\xi_{l+2} t}+e^{i\xi_{l-2} t}) |P_{\mathrm{vc}}| J_l(\eta_1,\eta_2,\eta_3,\eta_4) \right],
\end{gather}
with $\xi_l=\mathbf{k}^2/\mu+U_p+E_g+l\omega$.
Here we have assumed that the momentum matrix element is uniform in the $k$-space as,
\begin{equation}
    |P_{\mathrm{vc}}|^2 \sim \frac{E_g}{4\mu}.
\end{equation}
Using polar coordinates, the total transition probability is expressed as,
\begin{gather}
    W = \frac{|P_{\mathrm{vc}}|^2}{32\pi} \iint \sum_l \left| A_1(J_{l+1}(\eta_1,\eta_2,\eta_3,\eta_4) \right. \notag \\
    \left. +J_{l-1}(\eta_1,\eta_2,\eta_3,\eta_4)) \right. \notag \\
    \left. + A_2(J_{l+2}(\eta_1,\eta_2,\eta_3,\eta_4) \right. \notag \\
    \left. + J_{l-2}(\eta_1,\eta_2,\eta_3,\eta_4)) \right|^2 \delta (\xi_l) k^2 \sin{\theta} dk d\theta \label{W},
\end{gather}
where $\theta$ denotes the angle between the laser polarization direction, or polar angle, and $k=|\mathbf{k}|$.
If we put $\kappa_l=l\omega-E_g-U_p$ and note that $k=\sqrt{2\mu\kappa_l}$, we can rewrite $W$ as, 
\begin{gather}
    W = \frac{\pi^2|P_{\mathrm{vc}}|^2\mu^{3/2}}{4\sqrt{2}} \int_{0}^{\pi} d\theta \sum_l \left| A_1(J_{l+1}(\eta_1,\eta_2,\eta_3,\eta_4) \right. \notag \\
    \left. +J_{l-1}(\eta_1,\eta_2,\eta_3,\eta_4)) \right. \notag \\
    \left. + A_2(J_{l+2}(\eta_1,\eta_2,\eta_3,\eta_4) \right. \notag \\
    \left. + J_{l-2}(\eta_1,\eta_2,\eta_3,\eta_4)) \right|^2 \sqrt{\kappa_l} \sin{\theta}.
\end{gather}
The first two terms $J_{l\pm1}$ and the other two $J_{l\pm2}$ correspond to the individual contributions of the $\omega$ and $2\omega$ components, respectively. Their cross terms are nothing but a two-color mixing effect.

\subsubsection{Non-zero relative phase $\phi\neq0$}
Next, turn to the $\phi\neq 0$ case in Eq.~(\ref{vector_potential}).
The phase factor [Eq.~(\ref{phase_exp})] is modified as,
\begin{gather}
    \exp(i\int^t \Delta^{\mathbf{k}+\mathbf{A}(t')}_{\mathrm{vc}} dt') \notag \\
    =\sum_l \exp(i\left[\frac{k^2}{2\mu} + U_p + E_g + l\omega \right] t) \notag \\
    \times J_l^\phi(\eta'_1,\eta'_2,\eta'_3,\eta'_4,\eta'_5,\eta'_6,\eta'_7,\eta'_8),
\end{gather}
where $J_l^\phi$ is the $l$-th order ``twisted" multivariable Bessel function defined as,
\begin{gather}
    J_l^\phi(\eta'_1,\eta'_2,\eta'_3,\eta'_4,\eta'_5,\eta'_6,\eta'_7,\eta'_8) \notag \\
    = \sum_{l'} i^{l'} J_{l-l'}(\eta'_1,\eta'_2,\eta'_3,\eta'_4)\Tilde{J}_{l'}(\eta'_5,\eta'_6,\eta'_7,\eta'_8),
\end{gather}
where $\Tilde{J}_{l'}$ is defined as,
\begin{gather}
    \Tilde{J}_{l'}(\eta'_5,\eta'_6,\eta'_7,\eta'_8) \notag \\
    = \sum_{l_6,l_7,l_8} i^{-l_6-2l_7-3l_8} J_{l'-2l_6-3l_7-4l_8}(\eta'_5)I_{l_6}(\eta'_6)I_{l_7}(\eta'_7)I_{l_8}(\eta'_8),
\end{gather}
and,
\begin{gather}
    \eta'_1=\frac{A_1k\cos{\theta}}{\mu\omega}+\frac{A_1A_2}{2\mu\omega}\cos{\phi} \\
    \eta'_2=\frac{A_1^2}{8\mu\omega}+\frac{A_2}{2\mu\omega}\cos{\phi} \\
    \eta'_3=\frac{A_1A_2}{6\mu\omega}\cos{\phi}, 
    \eta'_4=\frac{A_2^2}{16\mu\omega}\cos{2\phi} \\
    \eta'_5=\frac{A_1A_2}{2\mu\omega}\sin{\phi}, 
    \eta'_6=\frac{A_2k\cos{\theta}}{2\mu\omega}\sin{\phi} \\
    \eta'_7=\frac{A_1A_2}{6\mu\omega}\sin{\phi}, 
    \eta'_8=\frac{A_2^2}{16\mu\omega}\sin{2\phi}
\end{gather}
Then, finally, we obtain the total transition probability as,
\begin{gather}
    W = \frac{|P_{\mathrm{vc}}|^2\mu^{3/2}}{16\sqrt{2}\pi} \notag \\
    \times \int_{0}^{\pi} d\theta \sum_l \left| A_1(J_{l+1}^\phi(\eta'_1,\eta'_2,\eta'_3,\eta'_4,\eta'_5,\eta'_6,\eta'_7,\eta'_8) \right. \notag \\
    \left. +J_{l-1}^\phi(\eta'_1,\eta'_2,\eta'_3,\eta'_4,\eta'_5,\eta'_6,\eta'_7,\eta'_8)) \right. \notag \\
    \left. + A_2(e^{i\phi}J_{l+2}^\phi(\eta'_1,\eta'_2,\eta'_3,\eta'_4,\eta'_5,\eta'_6,\eta'_7,\eta'_8) \right. \notag \\
    \left. + e^{-i\phi}J_{l-2}^\phi(\eta'_1,\eta'_2,\eta'_3,\eta'_4,\eta'_5,\eta'_6,\eta'_7,\eta'_8)) \right|^2 \sqrt{\kappa_l} \sin{\theta}. \label{Wphi}
\end{gather}
Again, the first two terms $J_{l\pm1}^\phi$ and the others $J_{l\pm2}^\phi$ correspond to excitation by $\omega$ and $2\omega$, respectively, whereas their cross-terms of them represent the two-color effect.

\section{\label{sec:results} Numerical assessments}

\subsection{\label{subsec:tddft} Time-dependent density functional theory}
To validate our analytical model, we perform first-principles calculations using the time-dependent density functional theory (TDDFT). The electron dynamics under a laser field are described by the time-dependent Kohn-Sham equation,
\begin{equation}
    i\hbar\frac{\partial}{\partial t} \psi_i(\mathbf{r},t) = h_{\mathrm{KS}}[n_e(\mathbf{r},t)] \psi_i(\mathbf{r},t),
\end{equation}
where the Kohn-Sham Hamiltonian under a vector potential $\mathbf{A}(t)$ in the velocity gauge is given by,
\begin{equation}
    h_{\mathrm{KS}}[n_e(\mathbf{r},t)] = \frac{1}{2m}[\mathbf{p}+e\mathbf{A}(t)]^2 + V_{\mathrm{eff}}[n_e(\mathbf{r},t)].
\end{equation}
Here, $n_e(\mathbf{r},t)$ is the time-dependent electron density, and $V_{\mathrm{eff}}$ consists of the ionic potential $V_{\mathrm{ion}}$, the Hartree potential $V_{\mathrm{H}}$, and the exchange-correlation potential $V_{\mathrm{xc}}$. The modified Becke-Johnson (mBJ) potential is used for $V_{\mathrm{xc}}$, ensuring accurate bandgap predictions \cite{Tran2009-ob}.

We consider a vector potential of the following form:
\begin{gather}
    \mathbf{A}(t) = f(t)(0,0,A_1 \cos{\omega t} + A_2 \cos{(2\omega t + \phi)}), \\
    f(t) = \left\{
        \begin{array}{ll}
        t/\tau & (0 \le t < \tau)\\
        1 & (\tau \le t < T-\tau)\\
        (T-t)/\tau & (T-\tau \le t < T)
        \end{array}
        \right.,
\end{gather}
where $A_1$ and $A_2$ denote amplitude factors for $\omega$ and $2\omega$, respectively, $T$  the foot-to-foot pulse width, and $\tau$ the rising time constant.

The instantaneous number of excited electrons is defined as,
\begin{equation}
    n_{\mathrm{ex}}(t) = \sum_{i\in \mathrm{CB}}2\left|\int d\mathbf{r}\sum_{j\in\mathrm{VB}} \psi_i^*(\mathbf{r},0)\psi_j(\mathbf{r},t)\right|^2
\end{equation}
We evaluate the time-averaged transition probability of an $N_e$ electron system per unit time and space as,
\begin{equation}
    W_{\mathrm{DFT}} = \frac{2(n_{\mathrm{ex}}(T-\tau)-n_{\mathrm{ex}}(\tau))}{(T-2\tau)N_e\Omega},
\end{equation}
where $\Omega$ stands for the simulation box volume.
The numerical calculation is executed using the SALMON code \cite{Noda2019-px}.

\subsection{\label{sec:comparison} Numerical comparison with TDDFT}
We assume that the fundamental photon energy is $\hbar\omega = 1.55$ eV, which corresponds to 800 nm wavelength. 
The $\omega$ and $2\omega$ fields are mixed with equal intensity.
We consider $\alpha$-SiO$_2$ crystal as a target.
The measured bandgap is ranging from 8.6 eV to 9.5 eV in spectroscopic studies \cite{Bart1992-pt}.
For the analytical expression, we assume $E_g=9$ eV and $\mu=0.28$ a.u. \cite{Otobe2019-cv}.
In our real-time TDDFT simulations, we set the laser pulse duration to $T=15$ fs with a gradual turn-on/off period of $\tau=3$ fs to reduce artifacts from sudden field application.
The unit cell considered is a cubic cell of $\alpha$-SiO$_2$ crystal containing 96 atoms, whose lattice constants are 
0.49,
0.85,
and 0.54 nm for the $x$, $y$, and $z$ axes, respectively.
The number of the spatial and $k$ grids are $26\times44\times38$ and $8\times8\times8$, respectively.
The time step is set to 0.004838 as. The calculated optical gap is 8.75 eV.

\subsubsection{Laser intensity dependence}
\begin{figure}[tb]
  \centering
  \includegraphics[keepaspectratio,width=0.95\hsize]{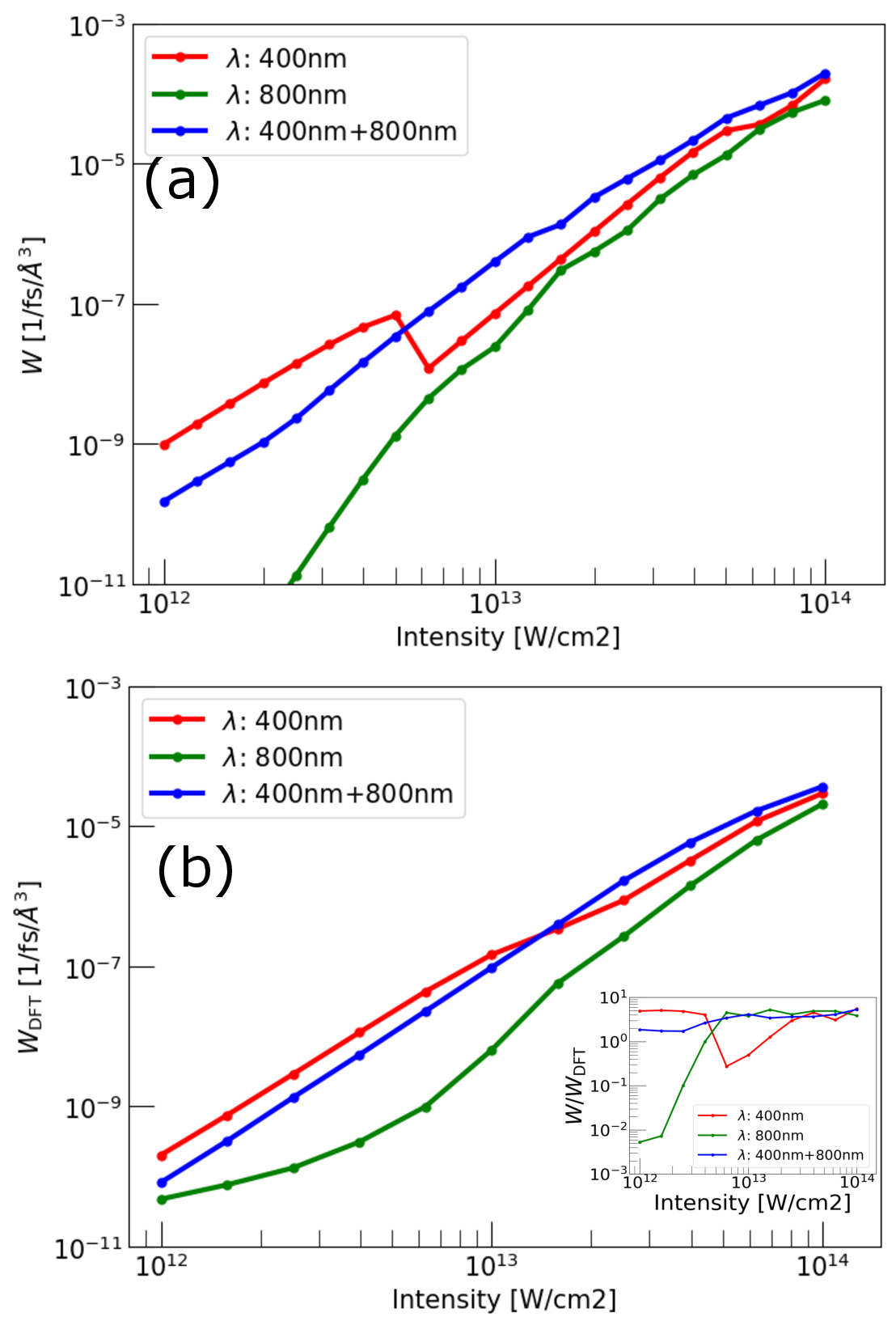}
  \caption[phase]{Intensity dependence of excitation probability calculated by the analytical formula (a) and TDDFT (b). Red line indicates $2\omega$ case, blue $\omega+2\omega$ case, and green  $\omega$ case. The inset of panel (b) is the intensity dependence of $W/W_{\mathrm{DFT}}$.}
  \label{intensity-dependence}
\end{figure}

Figure~\ref{intensity-dependence} illustrates the intensity dependence of the excitation probability for the $\phi=0$ case for monochromatic and two-color laser fields.
The results obtained by the analytical formula derived in the previous section [Fig.~\ref{intensity-dependence} (a)] reproduce the overall trend of the TDDFT results [Fig.~\ref{intensity-dependence} (b)] surprisingly well, albeit difference by a factor of approximately three [inset of Fig.~\ref{intensity-dependence} (b)].
For $I_{\mathrm{tot}}\gtrsim 10^{13} \ \mathrm{W/cm^2}$, the transition probability under two-color excitation is significantly enhanced compared to the single-color cases, while, at lower intensities, the $2\omega$ field yields the highest excitation probability.

In the perturbative regime, 6- and 3-photon absorption should be the leading order for 800~nm and 400~nm, respectively, which is indeed the case in the analytical results [Fig.~\ref{intensity-dependence}(a)].
In the low-intensity region, the TDDFT results suggest a process of a lower order than that of 6-photon absorption. 
This behavior indicates the one-photon absorption at the tail of the laser pulse in the frequency domain. Since the short pulse has a broad bandwidth, a weak tail above 9~eV induces one-photon process.
In the analytical formula, we assume the continuous wave which has a $\delta$-functional distribution in the frequency domain, leading to the $\delta$-function in Eq.~(\ref{W}).
The difference in the bandwidth of the applied field makes a large difference in 800~nm.

In the strong field regime, electrons are excited into higher energy bands where the system behaves more like a free-electron gas. As a result, the parabolic band assumption in our analytical model is a good approximation in this regime in all frequencies, leading to better agreement between our analytical model and TDDFT.

In Fig.~\ref{intensity-dependence} (a), we see a drop in excitation probability at $5\times10^{12} \ \mathrm{W/cm^2}$ for $400 \ \mathrm{nm}$ wavelength, originating from channel closing  due to ponderomotive increase in effective band gap \cite{Otobe2019-cv}. 
There are also signatures of channel closings at $5\times10^{13} \ \mathrm{W/cm^2}$ for $400 \ \mathrm{nm}$ wavelength, $1\times10^{13} \ \mathrm{W/cm^2}$ and $2\times10^{13} \ \mathrm{W/cm^2}$ for $800 \ \mathrm{nm}$ wavelength, and $1.5\times10^{13} \ \mathrm{W/cm^2}$ for $400 \ \mathrm{nm}+800 \ \mathrm{nm}$ cases. 
In contrast, in Fig.~\ref{intensity-dependence} (b), we do not see clear channel closings except at $2\times10^{13} \ \mathrm{W/cm^2}$ for wavelength of $400 \ \mathrm{nm}$ case.
This difference stems from the fact that pulsed laser fields used in the TDDFT simulations have finite spectral width.

\subsubsection{Relative phase dependence}
\begin{figure}[tb]
  \centering
  \includegraphics[keepaspectratio,width=0.95\hsize]{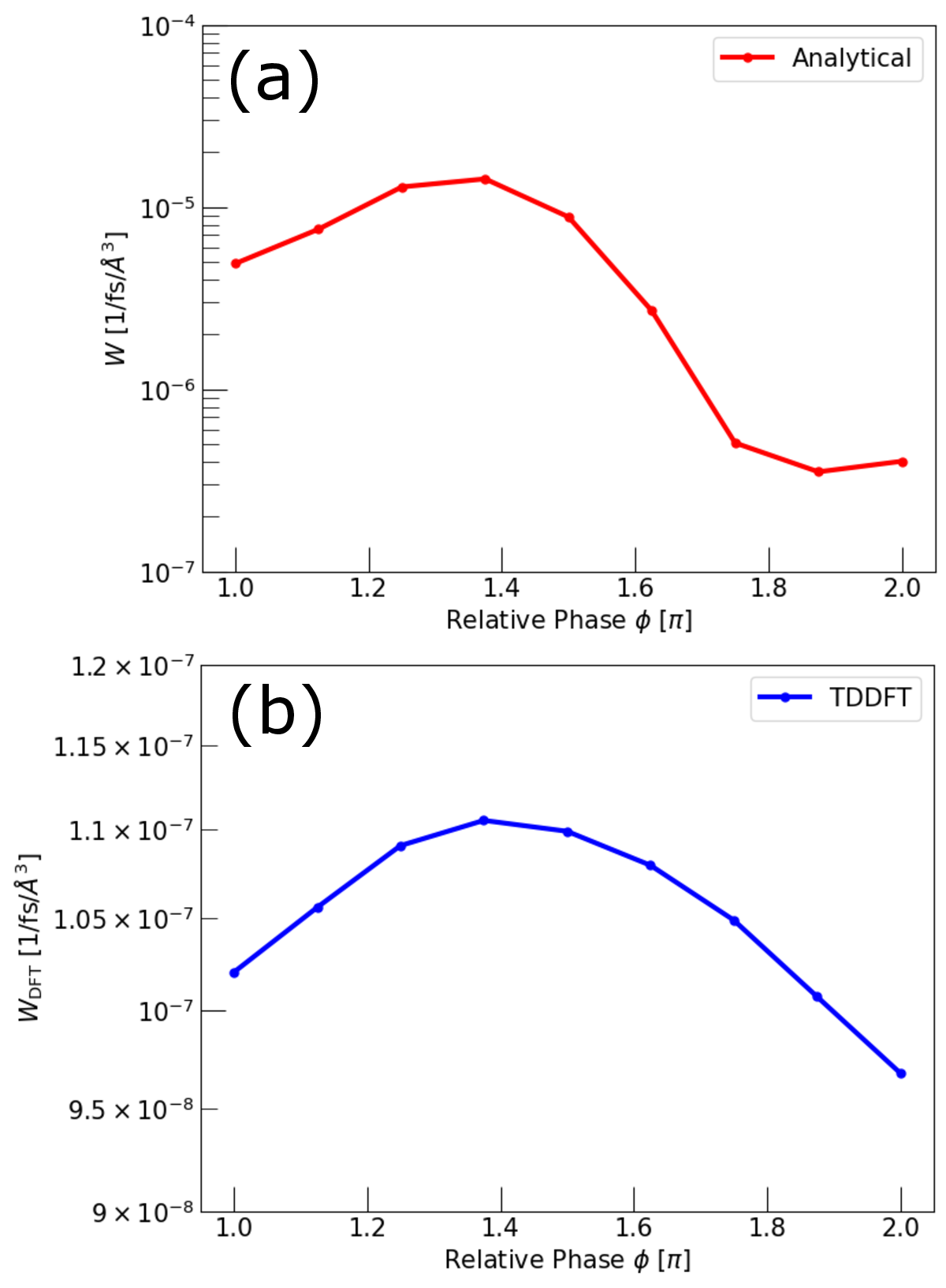}
  \caption[phase]{Relative-phase dependence of excitation probability calculated by the analytical formula (a) and TDDFT (b), with laser intensity $I_{tot}=10^{13} \ \mathrm{W/cm^2}$}
  \label{phase-dependence}
\end{figure}

Let us next discuss relative phase $\phi$ dependence of excitation probability for $I_{\mathrm{tot}} = 10^{13} \mathrm{W/cm^2}$ case.
Figure~\ref{phase-dependence} compares our analytical expression with TDDFT calculation results for an equally mixed two-color laser field.
Here, the domain of $\phi$ is set to $[\pi,2\pi]$. The vector potential waveform is invariant under the transformation $\phi \to 2\pi - \phi$ and $t \to -t$, so that the excitation probability is symmetric with respect to $\phi=\pi$ if the condition $|C_{\mathrm{vv}}^{\mathbf{k}}(t)| \sim 1$ holds.
We see the analytical formula and TDDFT calculation results in the same qualitative trend; excitation reaches minimum and maximum at $\phi\simeq \ 2\pi$ and $1.4\pi$, respectively.

\begin{figure}[tb]
  \centering
  \includegraphics[keepaspectratio,width=0.95\hsize]{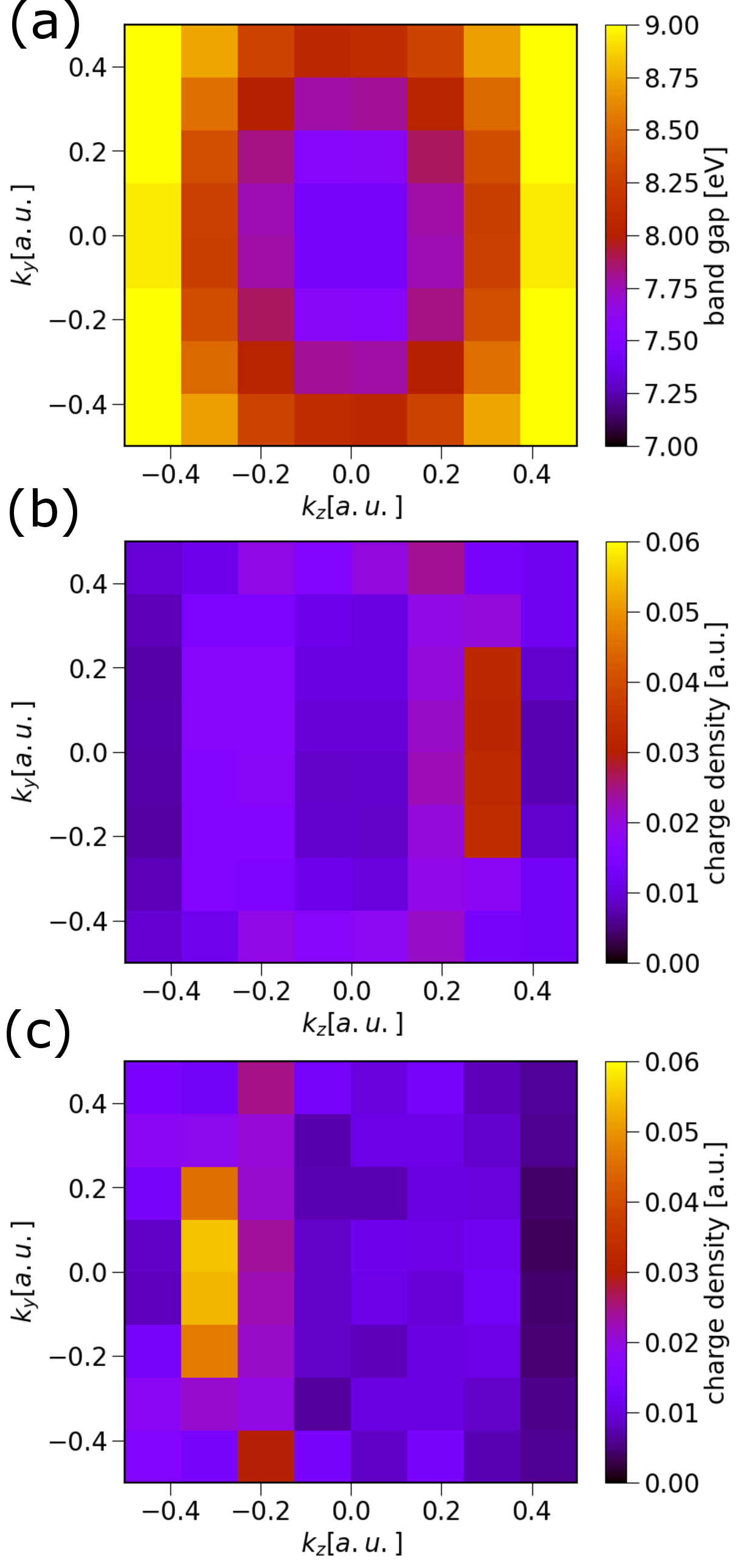}
  \caption[muz]{Bandgap at $k_x=0$ on the $yz$ plane within the first Brillouin zone is shown in pannel (a). Pannel (b) and (c) are carrier density map on the same plane with $\phi=1.4\pi$ and $\phi=2\pi$, respectively.}
  \label{muz}
\end{figure}

Since one of the major differences between our analytical model and TDDFT is the band structure landscape, we further analyze $k$-resolved excitation probabilities.
Figure~\ref{muz} shows the bandgap in the initial ground state and carrier distributions after laser irradiation on the $yz$-plane at $k_x=0$ calculated by DFT and TDDFT.
Although the energy bandgap shows a two-fold symmetry [Fig.~\ref{muz} (a)],
carrier density in Fig.~\ref{muz} (a) and (b) is largely anisotropic with respect to $k_z=0$, indicating non-uniform transition probability in the reciprocal space reflecting the anisotropy of the applied vector potentials. The carrier density integrated on this plane is 0.81 a.u. and 0.77 a.u. for Fig.~\ref{muz} (b) and (c), respectively. Although the local peak value is higher for $\phi=2\pi$, the total carrier excitation is higher for $\phi=1.4\pi$, where excitation is spread over a wide area of momentum.

\begin{figure}[tb]
  \centering
  \includegraphics[keepaspectratio,width=0.95\hsize]{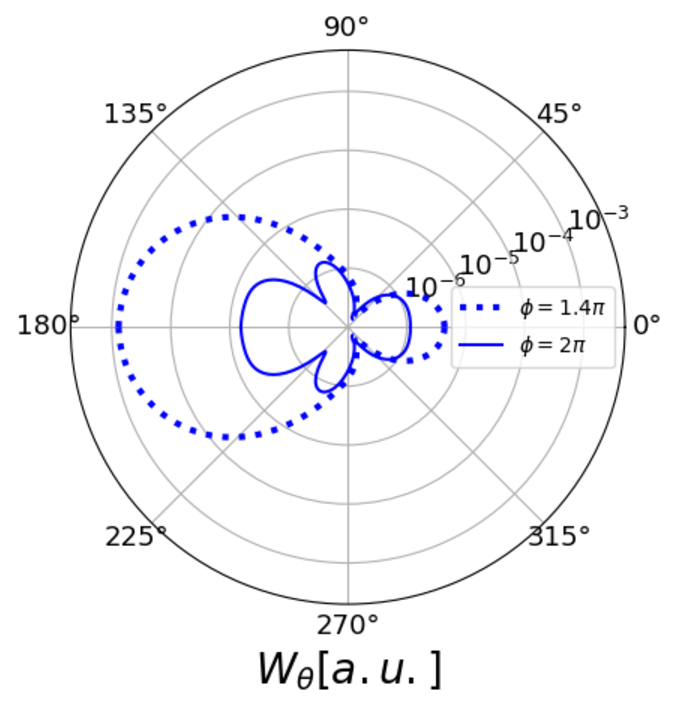}
  \caption[Wtheta]{Angle-resolved transition probability $W_{\theta}$ for $\phi=1.4\pi$ (dotted line) and $2.0\pi$ (solid line).}
  \label{Wtheta}
\end{figure}

Figure~\ref{Wtheta} shows the angle-resolved transition probability $W_{\theta}$, i.e., the integrand in Eq.~(\ref{Wphi}).
Comparing Fig.~\ref{muz}(b) and the dotted line in Fig.~\ref{Wtheta} for $\phi=1.4\pi$, we see the qualitative difference; the former shows much more population around $\theta =0^\circ$, while the latter reaches maximum at $\theta =180^\circ$.
A possible reason for the discrepancy is that the analytical two-band model focuses on excitations near $\Gamma$ point and assumes a parabolic dispersion without imposing periodicity in the $k$-space, while, in contrast, the TDDFT calculation respects the periodicity of the reciprocal lattice. 
As a result, the deviation between the two approaches may be more pronounced when the laser-field-driven electron motion in the
$k$-space exceeds the Brillouin zone boundaries.

For $\phi=2\pi$, on the other hand, we see relatively similar trends in Fig.~\ref{muz}(c) and solid line in Fig.~\ref{Wtheta}; there are three spikes, among which the most prominent one is in $\theta=180^\circ$, the second is in $\theta\simeq120^\circ$, and the third is in $\theta=0^\circ$.
Thus, although there is a quantitative difference between the two approaches stemming probably from the difference in band structure and periodicity,
they are surprisingly in excellent agreement in terms of the overall qualitative trends.

\section{\label{sec:conclusions} Conclusions}
We have derived an analytical expression for the ionization probability of dielectrics subjected to intense two-color laser fields, offering a compact and computationally efficient framework for describing electron excitation in strong-field regimes. Applying our formula to $\alpha$-quartz and benchmarking it against time-dependent density functional theory simulations, we have demonstrated that our model successfully captures the overall trends of ionization probability, particularly its dependence on laser intensity and relative phase.
The agreement between our analytical formula and TDDFT results supports the validity of the parabolic two-band approximation even in strong laser fields, where electronic transitions are dictated by band dispersion. The remaining quantitative discrepancies indicate that incorporating additional factors, such as band structure anisotropy and periodicity, could further improve accuracy. This work not only provides deeper insight into strong-field ionization dynamics but also establishes a practical theoretical tool for predicting laser-induced excitation processes in dielectrics.

\section*{\label{sec:acknowledgements} ACKNOWLEDGEMENTS}
This research was supported by MEXT Quantum Leap Flagship Program (MEXT Q-LEAP) Grant Number JPMXS0118067246 and JSPS KAKENHI Grant Number 24K23024 and 25K17367. 
M.T. gratefully acknowledges support from the Graduate School of Engineering, The University of Tokyo, Graduate Student Special Incentives Program. M.T. also gratefully thanks support from crowd funding platform  \it academist. \rm The TDDFT calculations are performed on the supercomputer Wisteria (the University of Tokyo).
\appendix
\def\thesection{\Alph{section}}
\section{\label{sec:appendixA}Details for phase factor Eq.~(\ref{phase_exp})}
In the phase factor transformation in Eq.~(\ref{phase_exp}), the arguments of exponential function is separated into time linear and oscillating parts with each frequency,
\onecolumngrid
\begin{gather}
    \exp(i\int^t \Delta^{\mathbf{k}+\mathbf{A}(t')}_{\mathrm{vc}} dt') = \exp \left(i\int^t \left[ \frac{k^2}{2\mu} + \frac{k\cos{\theta}}{\mu}(A_1 \cos{\omega t'} + A_2 \cos{2\omega t'}) \right. \right. \notag \\
    \left. \left. + \frac{1}{2\mu}(A_1^2 \cos^2{\omega t'} + A_2^2 \cos^2{2\omega t'} + 2A_1A_2\cos{\omega t'}\cos{2\omega t'}) + E_g \right] dt'\right) \notag \\
    = \exp \left(i\int^t \left[ \frac{k^2}{2\mu} + \frac{k\cos{\theta}}{\mu}(A_1 \cos{\omega t'} + A_2 \cos{2\omega t'}) \right. \right. \notag \\
    \left. \left. + \frac{1}{2\mu} \left\{ A_1^2 \frac{1+\cos{2\omega t'}}{2} + A_2^2 \frac{1+\cos{4\omega t'}}{2} + A_1A_2(\cos{3\omega t'} + \cos{\omega t'}) \right\} + E_g \right] dt'\right) \notag \\
    = \exp \left(i\int^t \left[ \frac{k^2}{2\mu} +\left(\frac{kA_1\cos{\theta}}{\mu}+\frac{A_1A_2}{2\mu} \right) \cos{\omega t'} + \left( \frac{kA_2\cos{\theta}}{\mu}+\frac{A_1^2}{4\mu} \right) \cos{2\omega t'} \right. \right. \notag \\
    \left. \left. + \frac{A_1A_2}{2\mu}\cos{3\omega t'} + \frac{A_2^2}{4\mu}\cos{4\omega t'} + \frac{A_1^2+A_2^2}{4\mu} + E_g\right] dt'\right) \notag
    \\
    = \exp \left(i \left[ \left(\frac{kA_1\cos{\theta}}{\mu\omega}+\frac{A_1A_2}{2\mu\omega} \right) \sin{\omega t} + \left( \frac{kA_2\cos{\theta}}{2\omega\mu}+\frac{A_1^2}{8\omega\mu} \right) \sin{2\omega t} \right. \right. \notag \\
    \left. \left. + \frac{A_1A_2}{6\omega\mu}\sin{3\omega t} + \frac{A_2^2}{16\omega\mu}\sin{4\omega t} \right] + i\left[ \frac{k^2}{2\mu} + \frac{A_1^2+A_2^2}{4\mu} + E_g \right] t\right) \notag \\
    =\exp(i\left[\frac{k^2}{2\mu} + \frac{A_1^2+A_2^2}{4\mu} + E_g \right] t) \sum_{l_1,l_2,l_3,l_4} \exp(i[l_1+2l_2+3l_3+4l_4]\omega t) J_{l_1}(\eta_1)J_{l_2}(\eta_2)J_{l_3}(\eta_3)J_{l_4}(\eta_4) \notag \\
    =\sum_l \exp(i\left[\frac{k^2}{2\mu} + \frac{A_1^2+A_2^2}{4\mu} + E_g + l\omega \right] t) J_l(\eta_1,\eta_2,\eta_3,\eta_4). \notag
\end{gather}
\twocolumngrid
\section{\label{sec:appendixB}On the transition probability Eq.~(\ref{W})}
The transition probability Eq.~(\ref{W}) is defined as $k$-space integral, transformed into the polar coordinate,
\begin{gather}
    W = \frac{\pi|P_{\mathrm{vc}}|^2}{16} \int_{\mathrm{BZ}} \sum_l \left| A_1(J_{l+1}(\eta_1,\eta_2,\eta_3,\eta_4) \right. \notag \\
    \left. +J_{l-1}(\eta_1,\eta_2,\eta_3,\eta_4)) \right. \notag \\
    \left. + A_2(J_{l+2}(\eta_1,\eta_2,\eta_3,\eta_4) \right. \notag \\
    \left. + J_{l-2}(\eta_1,\eta_2,\eta_3,\eta_4)) \right|^2 \delta (\xi_l) d\mathbf{k} \notag \\
    = \frac{\pi^2|P_{\mathrm{vc}}|^2}{8} \iint \sum_l \left| A_1(J_{l+1}(\eta_1,\eta_2,\eta_3,\eta_4) \right. \notag \\
    \left. +J_{l-1}(\eta_1,\eta_2,\eta_3,\eta_4)) \right. \notag \\
    \left. + A_2(J_{l+2}(\eta_1,\eta_2,\eta_3,\eta_4) \right. \notag \\
    \left. + J_{l-2}(\eta_1,\eta_2,\eta_3,\eta_4)) \right|^2 \delta (\xi_l) k^2 \sin{\theta} dk d\theta. \notag
\end{gather}

%


\end{document}